\documentclass[a4paper,fleqn]{cas-dc}
\usepackage{graphicx}
\usepackage{caption} 
\usepackage{float}
\usepackage{hyperref}
\hypersetup{
    colorlinks=true,
    urlcolor=black,     % This turns only URLs black
    citecolor=blue,     % Keeps citations blue (or change to black)
    linkcolor=blue      % Keeps internal links blue (or change to black)
}
\usepackage{cuted}
\usepackage{caption} % Required to make captions work outside of normal figure blocks
\usepackage[numbers]{natbib}
\def\tsc#1{\csdef{#1}{\textsc{\lowercase{#1}}\xspace}}
\tsc{WGM}
\tsc{QE}
\tsc{EP}
\tsc{PMS}
\tsc{BEC}
\tsc{DE}
\usepackage{graphicx}
\usepackage{cuted}
\usepackage[caption=false]{subfig} % The bulletproof fix for Elsevier cas-dc
\begin{document}
\let\WriteBookmarks\relax
\def\textpagefraction{.001}
\shorttitle{Spatial Neighboring Scattering Transform}
\shortauthors{M.T.A. Tawhid et~al.}

% --- TITLE CONFIGURATION ---
\title [mode = title]{Spatial Neighboring Scattering Transform: A Cross-Channel Amplitude Coupling Measure for EEG Connectivity}

\shortauthors{M.T.A. Tawhid et~al.}

% --- AUTHOR 1 ---
\author[1]{Md. Taksimul Ahsan Tawhid}[orcid=0009-0002-3530-3290] % <--- Inside separate square brackets right after the name
\cormark[1] 
\fnmark[1]  
\ead{ahtawhid12345@gmail.com}

% --- AUTHOR 2 ---

\author[1]{Nasif Ahmed Rafe}[orcid=0009-0006-7223-318X]

\cormark[1] % Triggers the corresponding author mark (*)

\fnmark[1]  % Triggers the equal contribution mark (†)

\ead{nasifrafe229@gmail.com}

% --- AUTHOR 3 ---

\author[2]{Alif Tahmid Priyom}[orcid=0009-0008-4491-525X] % Changed to 1 to match the shared affiliation

\ead{tahmidpriyom@gmail.com} 

% --- AUTHOR 3 ---

\author[3]{K. M. Mustafizur Rahman} [orcid=0000-0001-8739-2470]% Changed to 1 to match the shared affiliation

\ead{ mustafizbcc@eece.mist.ac.bd}

% --- UNIFIED AFFILIATION ---
\affiliation{organization={Department of EECE, Military Institute of Science and Technology (MIST)},
                addressline={Mirpur Cantonment},
                city={Dhaka},
                postcode={1216},
                country={Bangladesh}}

% --- FOOTNOTES ---
%\cortext[cor1]{Corresponding author(s).}
\fntext[fn1]{These authors contributed equally to this work.}
\begin{abstract}
The functional organization of the brain relies on coordinated activity 
across spatially distributed regions, making the analysis of inter-regional 
dependencies a fundamental problem in neuroscience and clinical EEG research. 
Existing connectivity measures address this predominantly through phase 
synchronization, which is vulnerable to volume conduction artifacts and 
discards amplitude-domain coupling that may independently reflect 
inter-regional interaction. This study introduces the Spatial Neighboring 
Scattering Transform (SNST), which extends the wavelet scattering transform 
to the multichannel setting by replacing the single-channel modulus with a 
cross-channel conjugate product, yielding two descriptors that jointly 
capture amplitude-envelope coupling between channels and its modulation 
across frequency scales. SNST was evaluated on the BCI Competition IV-2a 
motor-imagery dataset using a bias-corrected, false-discovery-rate-controlled 
statistical pipeline, with the validation criterion defined as spatial 
consistency of significant coupling across subjects for each imagery 
condition. The first-order descriptor identified statistically significant 
amplitude coupling within a central-parietal electrode neighborhood, 
reproduced consistently across all subjects and both imagery conditions. 
The second-order descriptor revealed that this coupling is periodically 
gated by slow rhythms, indicating a cross-frequency amplitude-modulation 
structure absent from single-frequency connectivity measures. Phase lag 
index and weighted phase lag index, computed under an identical correction 
procedure and verified robust to volume conduction, identified negligible 
significant coupling with zero overlap with SNST findings, demonstrating 
that amplitude-envelope coupling constitutes a largely distinct connectivity 
signal. These results establish SNST as a cross-channel scattering-based 
connectivity descriptor that recovers amplitude-envelope and cross-frequency 
coupling structure systematically inaccessible to phase-synchronization 
measures, applicable to any multichannel EEG analysis where amplitude-domain 
inter-regional dependence is of interest.
\end{abstract}
%\begin{graphicalabstract}====
    %\centering===
    %\includegraphics[width=\textwidth]{figs/GA-2.png}====
%\end{graphicalabstract}====

%\item SNST extends the wavelet scattering transform to the multichannel setting.

%\item The first-order descriptor identifies 
%statistically robust amplitude coupling.

%\item The second-order descriptor reveals that this 
%inter-regional coupling is periodically gated by slow rhythms.

%\item Phase lag index and weighted phase lag index, tested under an identical correction, demonstrate that amplitude-envelope coupling constitutes a largely distinct connectivity signal.

%\item SNST recovers inter-regional amplitude-domain connectivity structure systematically inaccessible to phase-synchronization measures.

\begin{keywords}
Spatial Neighboring Scattering Transform, Wavelet scattering transform, amplitude-envelope coupling, EEG functional connectivity, cross-frequency coupling, motor imagery, phase lag index
\end{keywords}

\maketitle

\section{Introduction}
Brain function is not localized to isolated regions but emerges from 
coordinated interaction among spatially distributed areas, with the activity 
of one region routinely conditioned on the state of its neighbors \cite{b1}. 
Evidence from cognitive neuroscience has progressively shifted away from the 
view that individual brain areas operate independently, instead supporting 
the view that cognition emerges from the dynamic interplay of distributed 
networks \cite{b2}. Network-level analyses of these interactions have 
revealed organizational properties, including small-world topology and hub 
structure, that recur across imaging modalities and are thought to underlie 
efficient information integration \cite{b3}. The same coupling that supports 
normal function also constrains pathology: large-scale network alterations, 
rather than damage confined to a single site, are now recognized as a 
defining feature of numerous neurological disorders \cite{b4}. This principle 
is well documented in electroencephalography (EEG) connectivity analyses, 
which have shown that local dysfunction in epilepsy, Alzheimer's disease, 
and Parkinson's disease propagates along existing functional pathways to 
alter activity in remote, ostensibly unaffected regions \cite{b5}. 
Characterizing how the activity of one region depends on another is therefore 
central to both basic neuroscience and clinical diagnosis.

Quantifying this inter-regional dependence has motivated decades of 
methodological development. Brain connectivity is commonly formalized at 
three levels---structural, functional, and effective---capturing, 
respectively, anatomical linkage, statistical dependence, and directed causal 
influence between regions \cite{b6}. Graph-theoretical measures built on 
these formalizations, including clustering coefficient, path length, and node 
centrality, were used to show that resting-state networks in Alzheimer's 
disease patients lose the small-world organization observed in healthy 
controls, directly linking network topology to clinical status \cite{b7}. To 
recover the directionality implicit in effective connectivity, Granger 
causality and its frequency-domain extension, partial directed coherence, 
were developed to infer which region's activity precedes and statistically 
predicts another's from multivariate time series \cite{b8}. These approaches 
collectively established that inter-regional relationships are measurable, 
structured, and clinically informative, but most were developed for, or first validated on, hemodynamic or source-localized signals rather than scalp-level electrophysiology.

Translating these ideas to EEG has relied predominantly on phase-based 
measures. The phase-locking value was introduced to quantify how consistently 
the phase relationship between two signals is maintained across trials, 
providing the first practical, frequency-specific synchronization estimate 
for neuroelectric data \cite{b9}. Phase-amplitude cross-frequency coupling, 
exemplified by the coupling between neocortical theta phase and gamma-band 
amplitude, extended this synchronization logic across frequency bands rather 
than within a single one \cite{b10}. However, EEG connectivity estimated at 
the sensor level is confounded by volume conduction: spatially adjacent 
electrodes can register the same underlying source with near-zero time lag, 
inflating both coherence and phase-locking estimates for pairs that are not 
genuinely interacting \cite{b11}. The phase lag index and its weighted 
extension were introduced specifically to suppress this artifact by 
discounting near-zero phase relationships \cite{b11,b12}, but in doing so 
they also discard genuine zero-lag and amplitude-only interactions, and none 
of these measures characterize coupling expressed through co-modulation of 
signal amplitude rather than phase.

The wavelet scattering transform (WST) offers a complementary, 
signal-processing-native alternative: a cascade of wavelet convolutions and 
modulus nonlinearities that yields stable, deformation-invariant 
time--frequency representations without requiring a learned model \cite{b13}. 
Extending this framework to audio established that its second-order 
coefficients explicitly characterize amplitude-modulation structure that 
linear time--frequency methods discard \cite{b14}, a property with direct 
relevance to physiological signals whose information often resides in slowly 
varying envelopes rather than instantaneous phase. Existing biomedical 
applications of WST, including its first uses on EEG for schizophrenia and 
alcoholism classification, nonetheless restrict this modulus operation to 
individual channels, computing each channel's representation independently 
before any cross-channel comparison \cite{b15,b16}. Consequently, WST has so 
far characterized within-channel dynamics but not inter-regional coupling, 
despite cross-channel coupling being the central object of interest in 
connectivity analysis, including in motor imagery, where directed connectivity 
analyses have reported increased inter-regional coupling between hemispheres 
during imagined hand movement, in a hand-specific direction \cite{b17}. We 
address this gap by extending the scattering operator with a cross-channel 
conjugate product, replacing the single-channel modulus with a pairwise term 
that directly encodes amplitude-envelope coupling and its cross-frequency 
modulation between channels. We evaluate the resulting descriptors on the BCI 
Competition IV-2a motor-imagery dataset \cite{b18}, and benchmark them 
against the phase lag index and weighted phase lag index under matched 
statistical correction.

\section{Methodology}\label{sec2}

\subsection{Signal Preprocessing}
Raw EEG recordings were band-pass filtered between 0.5 and 45 Hz using a fourth-order Butterworth filter, removing slow drift and frequencies beyond the range relevant to sensorimotor rhythms while retaining the delta-through-gamma spectrum required for cross-frequency analysis \cite{b19}. 
A 50 Hz notch filter (Q = 30) was additionally applied where line-noise power exceeded three times the local spectral baseline \cite{b19}. Ocular artifacts were corrected by linear regression of the three recorded electrooculogram channels onto each EEG channel, with the EOG-predicted component subtracted to leave the EOG-uncorrelated residual, following the regression-based correction method of Schl\"{o}gl et al. \cite{b20}.

\begin{figure*}[t!] % Standard wide float pushed cleanly to the top of the page
    \centering
    \includegraphics[width=\textwidth]{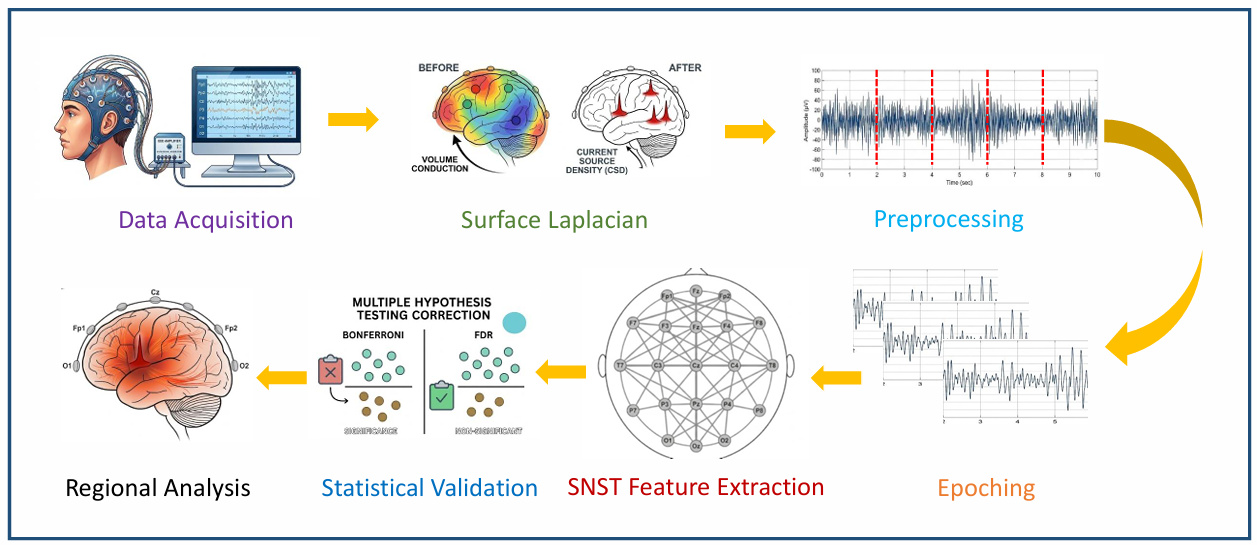} 
    \caption{Methodological flowchart of the overall EEG processing and analytical pipeline}
    \label{fig:figure_1}
\end{figure*}

A spherical-spline Surface Laplacian was subsequently applied to attenuate the spatial blurring introduced by volume conduction, retained only where it reduced mean pairwise 
inter-channel correlation for the majority of subjects \cite{b21}. Epochs were extracted relative to the cue-onset markers for left- and right-hand motor imagery, spanning the instructed imagery window (2--6 s post-cue, 4s duration, 1000 samples at 250 Hz), and trials flagged with the dataset's 
rejection marker were excluded \cite{b18}. Each preprocessing step was validated by direct, data-driven comparison of inter-channel correlation before and after its application, rather than adopted from precedent alone. The methodological workflow of the proposed method is illustrated in Figure~\ref{fig:figure_1}.

\subsection{Surface Laplacian}
Scalp-recorded EEG is an inherently smeared measurement of underlying 
cortical activity, volume conduction causes a focal source to register as 
a broad, gradually decaying potential across many electrodes, which inflates 
apparent connectivity between channels that are spatially close but not 
genuinely interacting \cite{b21,b22,b23}. The Surface Laplacian addresses 
this directly by computing the second spatial derivative of the scalp 
potential field, which acts as a spatial high-pass filter that suppresses 
the broad, smoothly varying component attributable to volume conduction while 
preserving sharp, locally focal activity, yielding a reference-free estimate 
of current source density at each electrode. The operator used here was 
constructed from the dataset's electrode coordinates with the spherical-spline 
method of Perrin et al. \cite{b24}, using spline order $m = 4$, regularization 
parameter $\lambda = 10^{-5}$, and a seven-term Legendre expansion 
\cite{b21,b22,b23,b24}.

\subsection{Spatial Neighboring Scattering Transform}
The wavelet scattering transform (WST) constructs stable,
time-shift-invariant signal representations by cascading wavelet
convolution with a pointwise modulus and a final low-pass averaging
operation \cite{b13,b25}. For a single-channel signal $x$, the
first-order WST coefficient at wavelet scale $\lambda$ is:
\begin{equation}
\label{eq:wst}
S_1^{\text{WST}}(t,\lambda) = \big( |x * \psi_\lambda| * \phi_J \big)(t)
\end{equation}
where $\psi_\lambda$ is a band-pass wavelet centered at frequency
$\lambda$, $\phi_J$ is a low-pass averaging window with invariance scale
$T = 2^J$ samples, and the outer parentheses indicate that the entire
convolution chain, not $\phi_J$ alone, is evaluated at time $t$.

SNST generalizes this construction to the multichannel setting by
replacing the single-channel modulus with a cross-channel conjugate
product. If $x_m$ and $x_n$ denote the recordings at two distinct
channels. SNST's first-order coefficient is defined as:
\begin{equation}
\label{eq:u1}
U_1(t;m,n,\lambda) = \big| (x_m * \psi_\lambda)(t) \cdot
(x_n * \psi_\lambda)^*(t) \big|
\end{equation}
\begin{equation}
\label{eq:s1}
S_1(m,n,\lambda) = \big( U_1(\cdot\,;m,n,\lambda) * \phi_J \big)(t)
\end{equation}
\begin{equation}
\label{eq:s1norm}
S_1^{\text{norm}}(m,n,\lambda) =
\frac{S_1(m,n,\lambda)}{\sqrt{E_m(\lambda)\,E_n(\lambda)}}
\end{equation}
where $(\cdot)^*$ denotes complex conjugation, $E_m(\lambda)$ and
$E_n(\lambda)$ are each channel's own band-limited energy at scale
$\lambda$, and Equation~(\ref{eq:s1norm}) is bounded in $[0,1]$ by the
Cauchy--Schwarz inequality. Here, $U_1$ is the instantaneous
cross-channel envelope product at frequency $\lambda$; its time average
$S_1$ quantifies the degree to which channels $m$ and $n$ co-modulate
in amplitude at that frequency, and $S_1^{\text{norm}}$ expresses this
quantity independently of either channel's absolute power.

\begin{figure*}[t!] % Standard wide float pushed cleanly to the top of the page
    \centering
    \includegraphics[width=\textwidth]{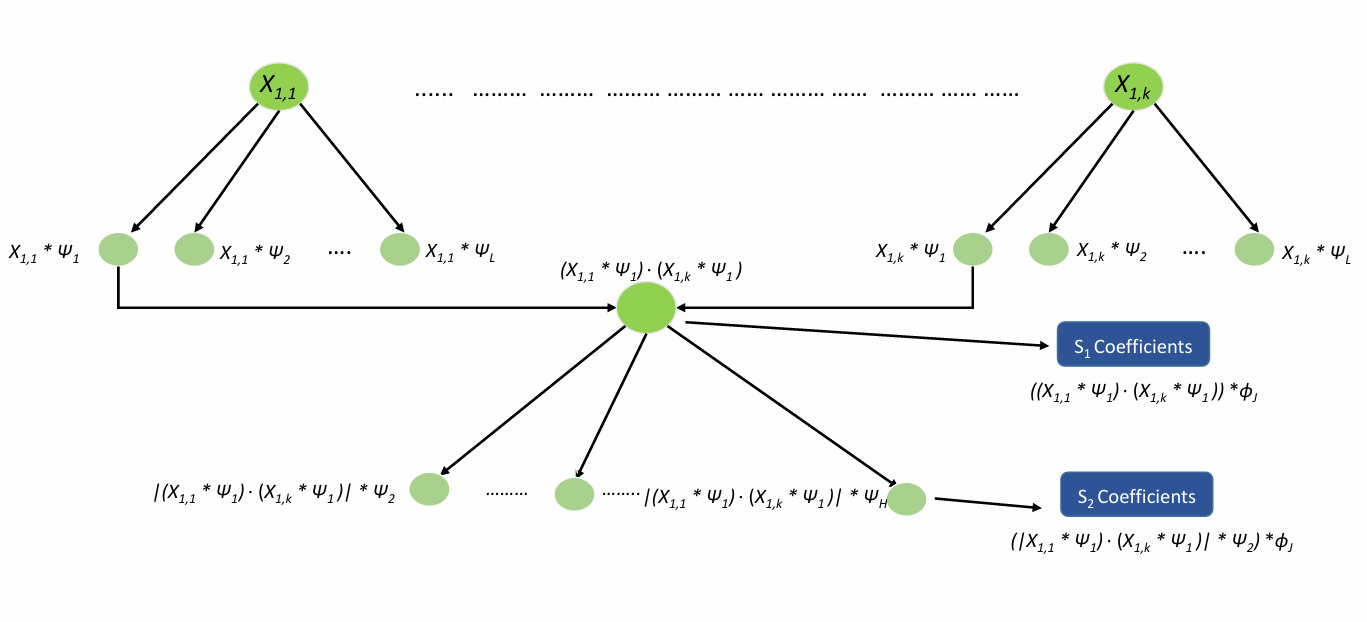}
    \caption{Schematic representation of multilevel coefficient extraction using the proposed Spatial Neighboring Scattering Transform (SNST). For the first epoch of channel-1, the coefficient will be extracted from 2nd,3rd,...upto channel-k with the same timed epoch.}
    \label{fig:Fig_2}
\end{figure*}

Unlike phase-synchronization measures, $S_1^{\text{norm}}$ is insensitive
to the relative phase between channels and instead captures envelope
co-fluctuation directly---a quantity that conventional coherence- and
phase-based connectivity measures do not estimate \cite{b26,b27}.

Because $U_1$ is itself a time-varying signal, it can be further
decomposed by a second wavelet filter bank operating at frequencies
strictly lower than $\lambda$, yielding the second-order coefficient:
\begin{equation}
\label{eq:s2}
S_2(m,n,\lambda,\lambda') = \big( \big(U_1(\cdot\,;m,n,\lambda) *
\psi_{\lambda'}\big) * \phi_J \big)(t), \qquad \lambda' < \lambda
\end{equation}
\begin{equation}
\label{eq:s2norm}
S_2^{\text{norm}}(m,n,\lambda,\lambda') =
\frac{S_2(m,n,\lambda,\lambda')}{\max\big(0.1\cdot
\widetilde{S}_1(m,n),\, \varepsilon\big)}
\end{equation}
where $\widetilde{S}_1(m,n)$ is the median of $S_1(m,n,\lambda)$ across
$\lambda$ for that pair and $\varepsilon$ is a numerical floor. The
floored denominator in Equation~(\ref{eq:s2norm}) was required because
$S_1$ is unbounded above zero and, unlike $S_1^{\text{norm}}$, has no
Cauchy--Schwarz ceiling; dividing by a per-pair value too close to zero
was found to inflate $S_2^{\text{norm}}$ to physically meaningless
magnitudes for pairs with weak baseline coupling. $S_2^{\text{norm}}$
characterizes whether the cross-channel envelope coupling at frequency
$\lambda$ is itself rhythmically modulated at a slower frequency
$\lambda'$. This is the cross-channel analogue of the amplitude-modulation
structure recovered by second-order WST coefficients in audio analysis
\cite{b14}, here applied to a coupling signal rather than to a single
channel's own envelope, and is conceptually related to, but mathematically
distinct from, phase-amplitude cross-frequency coupling \cite{b10}:
$S_2^{\text{norm}}$ characterizes modulation of an inter-channel coupling
strength, not modulation of one channel's amplitude by another's phase.

Both wavelet banks used Morlet wavelets as the mother wavelet, whose
analytic form directly yields the complex signal required for the
cross-channel conjugate product in Equations~(\ref{eq:u1}) and
(\ref{eq:s1}), and whose constant-$Q$ property ensures uniform relative
frequency resolution across scales. The first-order bank was configured
with $J_1 = 6$ octaves and $Q_1 = 2$ wavelets per octave, producing 14
center frequencies within the reliable passband, covering the full delta-through-beta EEG spectrum with
two filters per octave. The second-order bank used $J_2 = 6$ and
$Q_2 = 1$, providing a coarser decomposition appropriate for the slower
modulation frequencies targeted by $S_2^{\text{norm}}$. Both banks were
calibrated using the constant-$Q$, constant-bandwidth elbow construction
standard to scattering-transform implementations \cite{b14,b28}, with
$\phi_J$'s bandwidth set to $1/2^J$ samples by definition, so that all
numerical choices in the construction are derived from the $(J,Q)$
parameters rather than fitted post hoc. The multilevel feature extraction
of the proposed SNST method is shown in Figure~\ref{fig:Fig_2}.

\subsection{Statistical Validation and Spatial Consistency of SNST 
Coefficients}
Statistical significance of individual channel-pair coefficients was 
assessed with a one-sample $t$-test across subjects against the cross-pair 
baseline at the corresponding frequency, with Benjamini--Hochberg 
false-discovery-rate (FDR) correction applied across all tested pairs and 
frequencies to control for the large number of simultaneous comparisons 
inherent to a dense connectivity analysis \cite{b29}. The one-sample 
$t$-test was selected over non-parametric alternatives because the 
Wilcoxon signed-rank test, at $n = 9$ subjects across 3{,}234 simultaneous 
combinations, has a minimum achievable $p$-value of $0.0039$ that exceeds 
the Benjamini--Hochberg threshold for the top-ranked combination 
($\alpha/N = 1.55 \times 10^{-5}$), rendering it mathematically incapable 
of producing any significant result regardless of effect size. Synthetic 
epochs containing only independent noise were used throughout development 
to confirm that $S_1^{\text{norm}}$ and $S_2^{\text{norm}}$ produce no 
spurious significant coupling in the absence of true cross-channel 
dependence, and that planted coupling structures of specified frequency 
and spatial location are correctly recovered, before any coefficient was 
computed on real data.

To determine whether significant coupling was spatially organized rather 
than incidental, FDR-corrected coefficients were projected onto scalp 
topographies for a set of seed electrodes, using electrode positions from 
the dataset's standard montage \cite{b30} and topographic interpolation 
implemented in MNE-Python \cite{b31}. For each seed, a target electrode's 
coefficient was retained in the topography only if it survived FDR 
correction; the resulting maps were generated separately for each subject 
and as a group average, and a coupling pattern was classified as spatially 
consistent if the same neighboring electrodes survived correction for the 
same seed across the majority of subjects and in the corresponding 
group-averaged topography.

\subsection{Comparison Against Phase-Based Connectivity Measures}
Three established phase-synchronization estimators were computed from the 
same preprocessed epochs for comparison with $S_1^{\text{norm}}$: the 
phase-locking value (PLV)~\cite{b9}, the phase lag index 
(PLI)~\cite{b11}, and its weighted extension (wPLI)~\cite{b12}. For each 
measure, the analytic phase of every channel was obtained via the Hilbert 
transform after band-pass filtering into five frequency bands, and pairwise synchronization was computed per epoch and 
averaged within subject, mirroring the per-subject averaging applied to 
$S_1^{\text{norm}}$.

PLV quantifies the consistency of the inter-channel phase difference 
across time but is confounded by volume conduction, which produces a 
near-zero-lag phase relationship between spatially adjacent electrodes 
that PLV does not distinguish from genuine synchronization~\cite{b11}. 
PLI addresses this by retaining only the sign, rather than the magnitude, 
of the instantaneous phase difference, which discounts contributions 
clustered at zero lag while preserving consistently lagged 
interactions~\cite{b11}; wPLI further weights each contribution by its 
distance from zero lag, increasing sensitivity relative to PLI's binary 
criterion while retaining the same resistance to volume conduction 
\cite{b12}. The susceptibility of PLV and the resistance of PLI and wPLI 
to this artifact were each verified directly prior to their use, on 
synthetic epochs in which all channels shared a single underlying source 
at near-zero lag: PLV registered spuriously high coupling under this 
condition, whereas PLI and wPLI yielded near-chance false-positive rates.

All three phase-based measures were bias-corrected using the same 
cross-pair $z$-score normalization applied to $S_1^{\text{norm}}$ 
, with the per-subject mean computed after epoch averaging, 
before being tested for statistical significance using the identical 
pipeline: a one-sample $t$-test across subjects 
against the cross-pair baseline at each frequency band, with 
Benjamini--Hochberg false-discovery-rate correction applied across all 
tested combinations \cite{b29}.

\section{Results}\label{sec3}

\subsection{Preprocessing}
\begin{center}
    \nopagebreak
    \includegraphics[width=\linewidth]{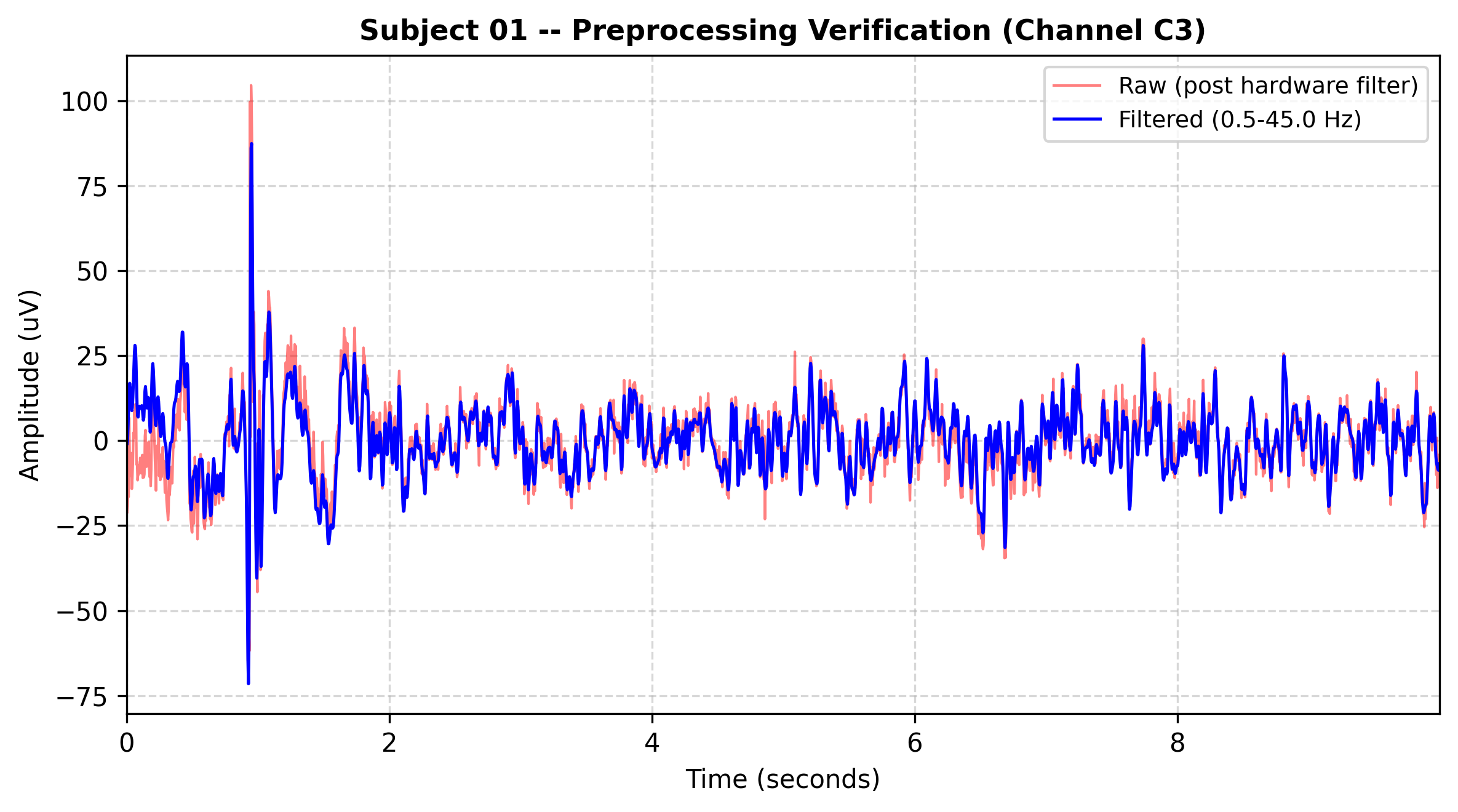}
    \captionof{figure}{Comparison of raw and filtered EEG signals for 
    Subject 01 (Channel C3)}
    \label{fig:Fig_3}
\end{center}
Linear regression of the three electrooculogram channels onto each EEG 
channel reduced mean pairwise inter-channel correlation from 0.981--0.995 
to 0.749--0.881 across the nine subjects, confirming removal of a 
substantial shared ocular component without eliminating genuine 
inter-channel structure entirely. Spectral inspection of the 50~Hz 
line-noise-to-neighbor-frequency power ratio identified residual 
contamination in one of nine subjects (ratio 4.53, against a clean range 
of 0.22--2.94 for the remaining eight); a notch filter was applied to 
this subject only. The comparison of preprocessed and raw signals is shown in 
Figure~\ref{fig:Fig_3}.

The Surface Laplacian reduced mean pairwise inter-channel correlation in 
all nine subjects, from a pre-Laplacian range of 0.726--0.870 to a 
post-Laplacian range of 0.249--0.296 (mean reduction 65.9\%, range 
59.5--71.0\%). This magnitude of reduction is consistent with prior 
quantitative analyses of volume conduction in scalp EEG, which report 
that electrodes separated by 10--12~cm can show appreciable spurious 
correlation from volume conduction alone, substantially attenuated by 
Laplacian filtering \cite{b32,b33}. The mitigation of spatial volume 
conduction artifacts and the resulting adjustments in cross-channel 
connectivity are shown in Figure~\ref{fig:Fig_4,5}. Because all nine subjects showed improvement here, exceeding the 
pre-specified 70\%-of-subjects adoption threshold, the 
Laplacian-referenced signal was retained for all subsequent analysis. 
Epoching around the cue-onset markers, restricted to the 2--6~s imagery 
window and excluding trials flagged by the dataset's own artifact 
marker\cite{b18}.

\begin{figure*}[t!]
    \centering
    \includegraphics[width=0.9\textwidth,height=4.5cm,keepaspectratio]
    {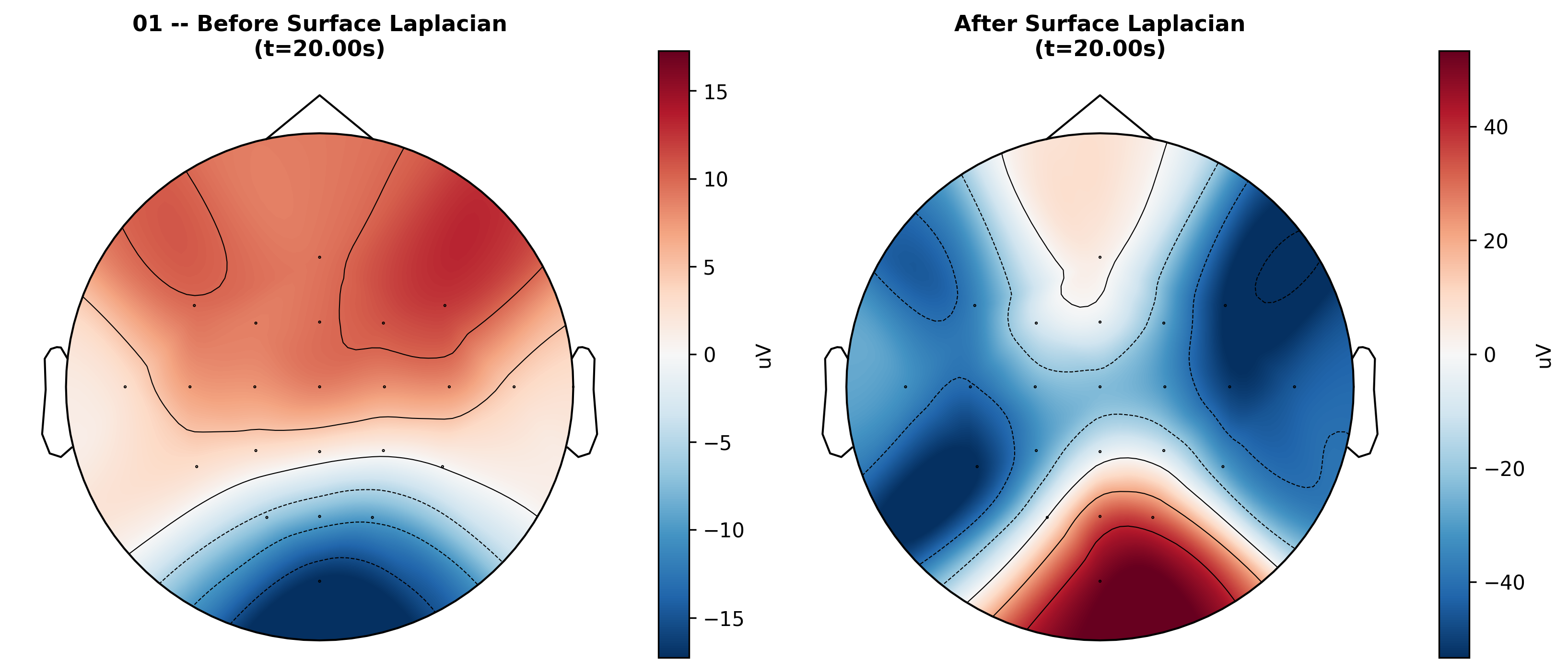} \\
    \vspace{2pt}
    {\small (a) The spatial distribution of scalp potentials is captured 
    at an instantaneous time point} \\ \vspace{15pt}
    \includegraphics[width=0.85\textwidth,height=4.5cm,keepaspectratio]
    {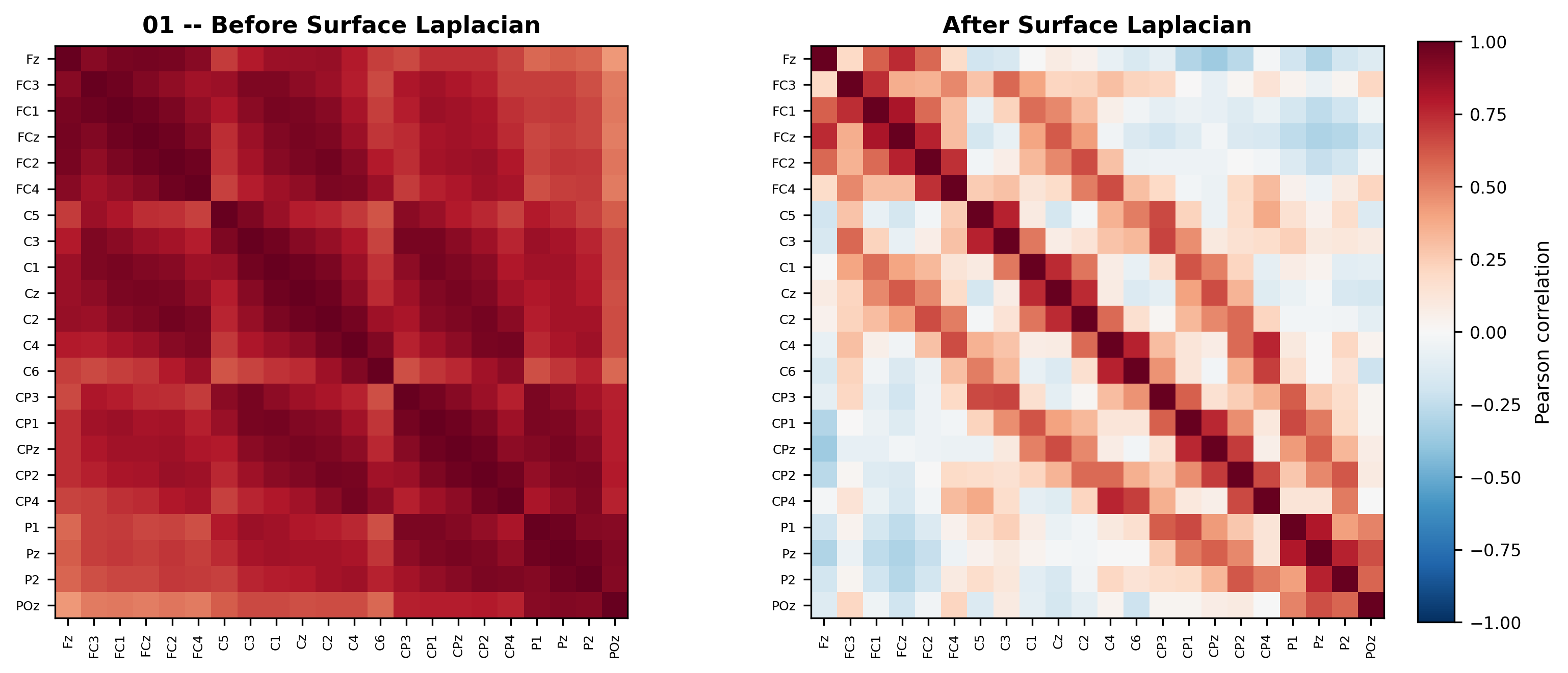} \\
    \vspace{2pt}
    {\small (b) The heatmaps evaluate the linear dependencies across the 
    multi-channel electrode array} \\ \vspace{10pt}
    \caption{Impact of the Surface Laplacian transform on EEG spatial 
    resolution and cross-channel connectivity}
    \label{fig:Fig_4,5}
\end{figure*}

\subsection{$S_1^{\text{norm}}$ Reveals Consistent Amplitude-Envelope 
Coupling Localized to a Central-Parietal Neighborhood}

A one-sample $t$-test against the cross-pair baseline, with 
Benjamini--Hochberg false-discovery-rate (FDR) correction applied across 
all tested (pair, frequency) combinations, identified 306 
combinations as significant for left-hand imagery (9.46\%) and 365 for 
right-hand imagery (11.29\%). The single most consistent effect across 
both conditions was C4--CP4, which survived correction at multiple 
frequencies in left-hand imagery (mean $z = 0.668$ at 32.65~Hz, 
$t = 21.09$, $p < 0.00001$, all subjects sharing the same sign) and 
recurred across nine of the fourteen tested frequencies in right-hand 
imagery, each with all sign agreement. A cluster of anatomically 
adjacent pairs---C5--C3, C3--CP3, C1--CP1, C4--C6, Pz--P2, P1--Pz, 
and FC1--FCz---recurred among the strongest combinations in both 
conditions, indicating that significant coupling was organized within a 
local centro-parietal neighborhood rather than confined to a single 
isolated pair. 
\begin{figure*}[t!]
    \centering
    \includegraphics[width=0.85\textwidth,height=3.5cm,keepaspectratio]
    {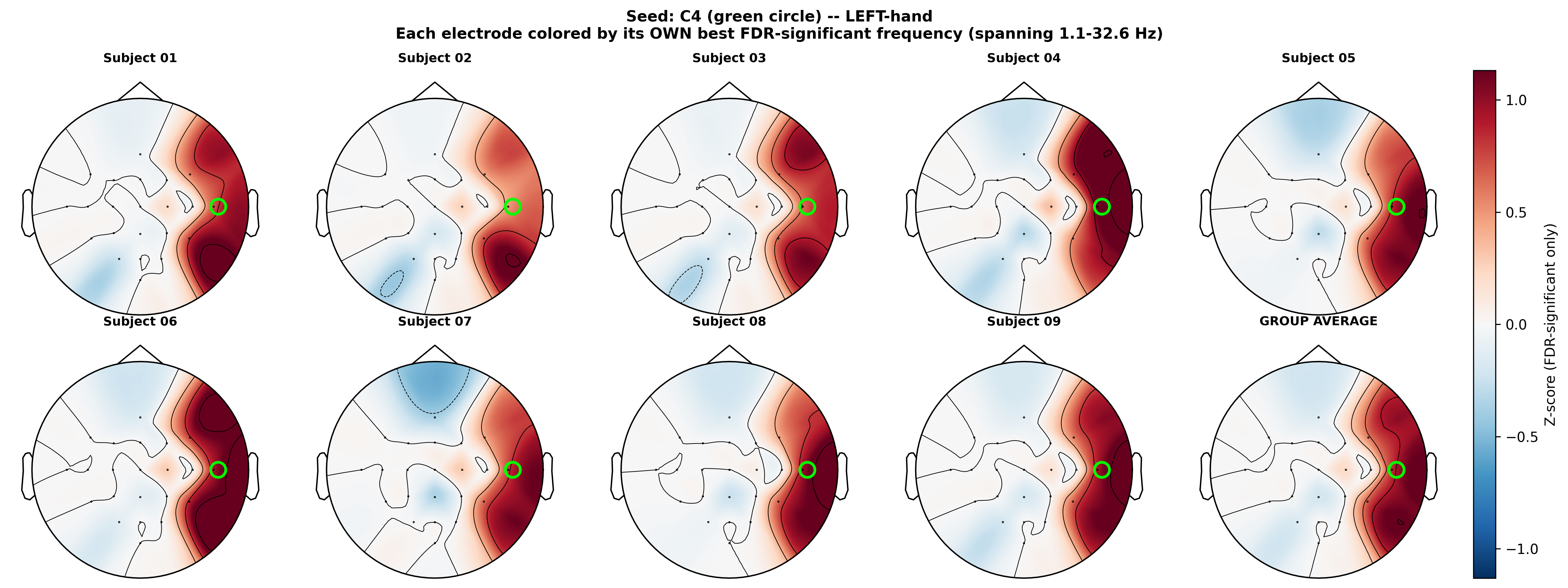} \\
    \vspace{2pt}
    {\small (a) Seed-based topography to show Amplitude envelope modulation for C4 seed during left-hand motor imagery} \\ \vspace{15pt}
    \includegraphics[width=0.85\textwidth,height=3.5cm,keepaspectratio]
    {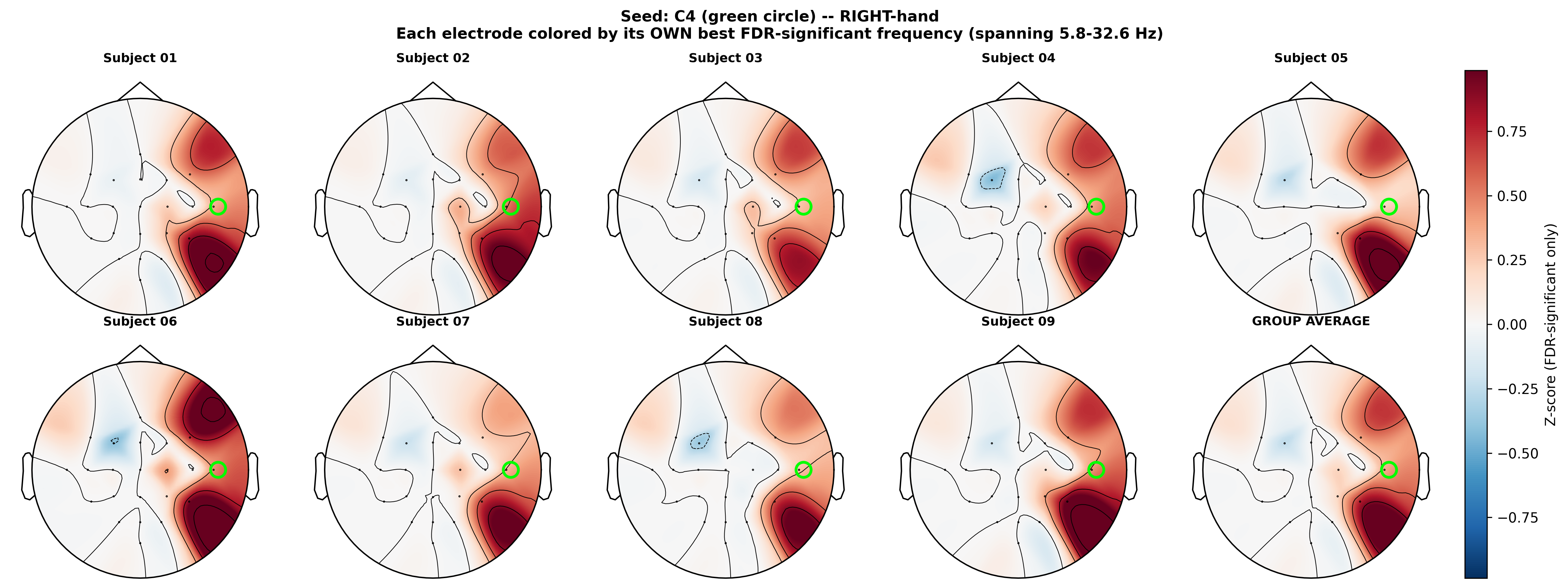} \\
    \vspace{2pt}
    {\small (b) Seed-based topography to show Amplitude envelope modulation for C4 seed during right-hand motor imagery} \\ \vspace{15pt}
    \includegraphics[width=0.85\textwidth,height=3.5cm,keepaspectratio]
    {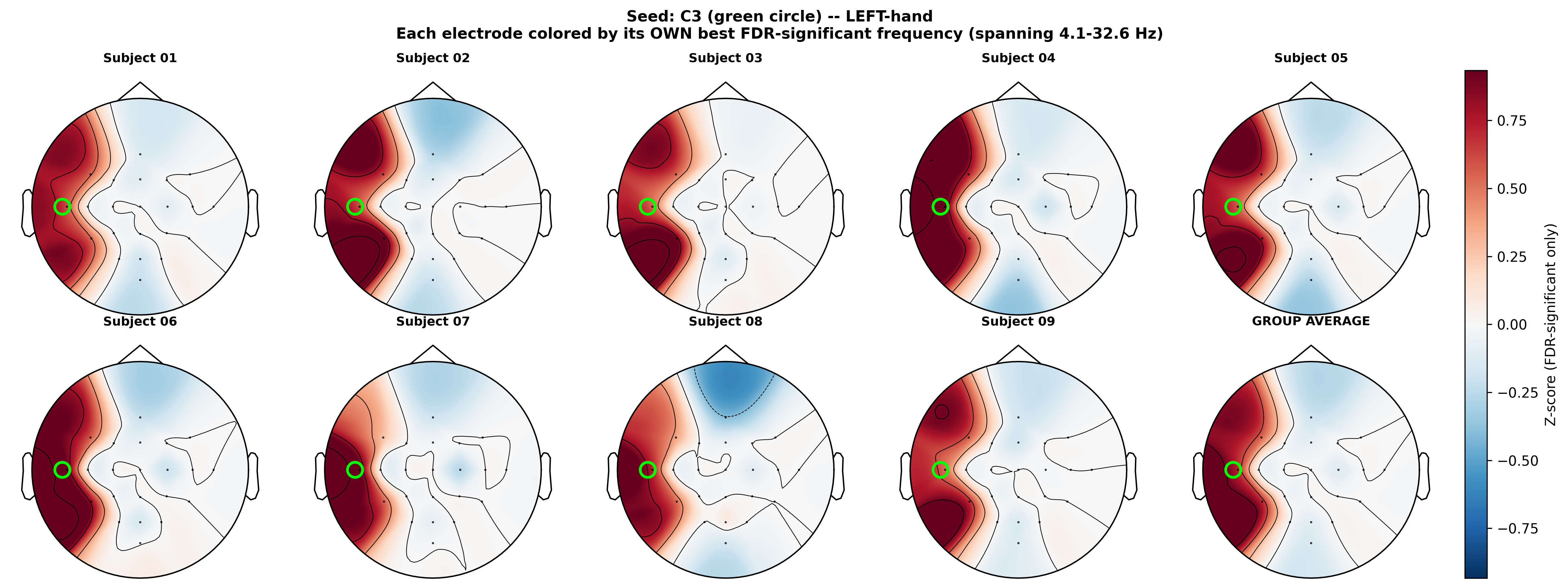} \\
    \vspace{2pt}
    {\small (c) Seed-based topography to show Amplitude envelope modulation for C3 seed during left-hand motor imagery} \\ \vspace{15pt}
    \includegraphics[width=0.85\textwidth,height=3.5cm,keepaspectratio]
    {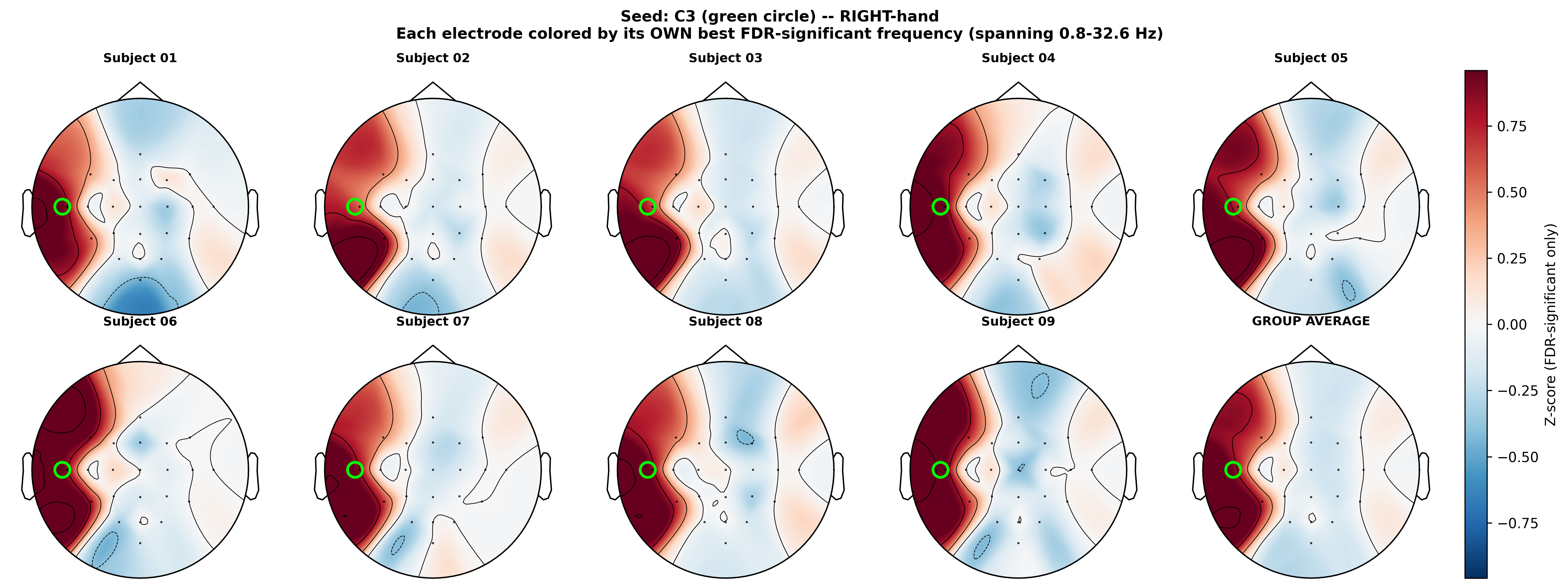} \\
    \vspace{2pt}
    {\small (d) Seed-based topography to show Amplitude envelope modulation for C3 seed during Right-hand motor imagery} \\ \vspace{15pt}
    \caption{Topographic visualization of amplitude envelope modulation and neighboring channel influence for C3 and C4 seed configurations during left- and right-hand motor imagery.}
    \label{fig:S1}
\end{figure*}

Projecting FDR-significant $S_1^{\text{norm}}$ coefficients onto scalp 
topographies for two seed electrodes showed that this regional 
structure was spatially organized rather than incidental to the aggregate 
statistic: for the C4 seed, target electrodes surviving correction were 
concentrated at FC4, C6, CP4, and CP2---electrodes immediately adjacent 
to C4---in 7 of 21 (left-hand) and 6 of 21 (right-hand) channels. The 
C3 seed showed a comparable or broader pattern ($8/21$ left-hand, 
$15/21$ right-hand), and the Cz seed showed the most extensive 
involvement of any seed ($14/21$ left-hand, $10/21$ right-hand). The amplitude envelope modulation of neighboring channels relative to the selected seed regions is illustrated in Figure~\ref{fig:S1}. Specifically, the topographic maps map the spatial propagation and localized effects across the scalp for both C3 and C4 seed electrodes during distinct left- and right-hand motor imagery tasks. In 
every case, the same neighboring electrodes survived correction 
consistently across the per-subject topographies and the corresponding 
group-averaged map, indicating that the coupling detected by 
$S_1^{\text{norm}}$ reflects a spatially coherent regional pattern 
shared across subjects rather than a result driven by a small subset of 
individuals. This localization to the C4--CP4 neighborhood and 
surrounding central-parietal region is consistent with prior 
connectivity analyses on this dataset that identify these electrodes as 
the primary locus of motor-imagery-related inter-channel interaction 
\cite{b34,b35}.

\subsection{$S_2^{\text{norm}}$ Identifies Slow-Rhythm Gating of 
Cross-Frequency Envelope Coupling}
Applying the breadth-of-modulation criterion to $S_2^{\text{norm}}$---the 
proportion of candidate fast-carrier frequencies showing elevated coupling 
for a given slow frequency identified 396 of 
1,386 tested (pair, slow-frequency) combinations as significant for 
left-hand imagery (28.57\%) and 387 of 1,386 for right-hand imagery 
(27.92\%), with 211 combinations shared between conditions. The broadest 
modulation in either condition was observed at C4--C6 (breadth fraction 
$0.76$ at $5.47$~Hz, left-hand; $0.63$ at the same frequency, right-hand), 
with C4--CP4 also reaching high breadth ($0.56$ left-hand, $0.61$ 
right-hand). The slow frequencies producing the broadest modulation were 
concentrated below $11$~Hz in both conditions, a range consistent with the established 
principle that slow amplitude co-fluctuations constitute a key mechanism 
constraining and gating faster inter-regional network activity 
\cite{b36,b37}.

\begin{figure*}[t!]
    \centering
    \includegraphics[width=0.85\textwidth,height=3.5cm,keepaspectratio]
    {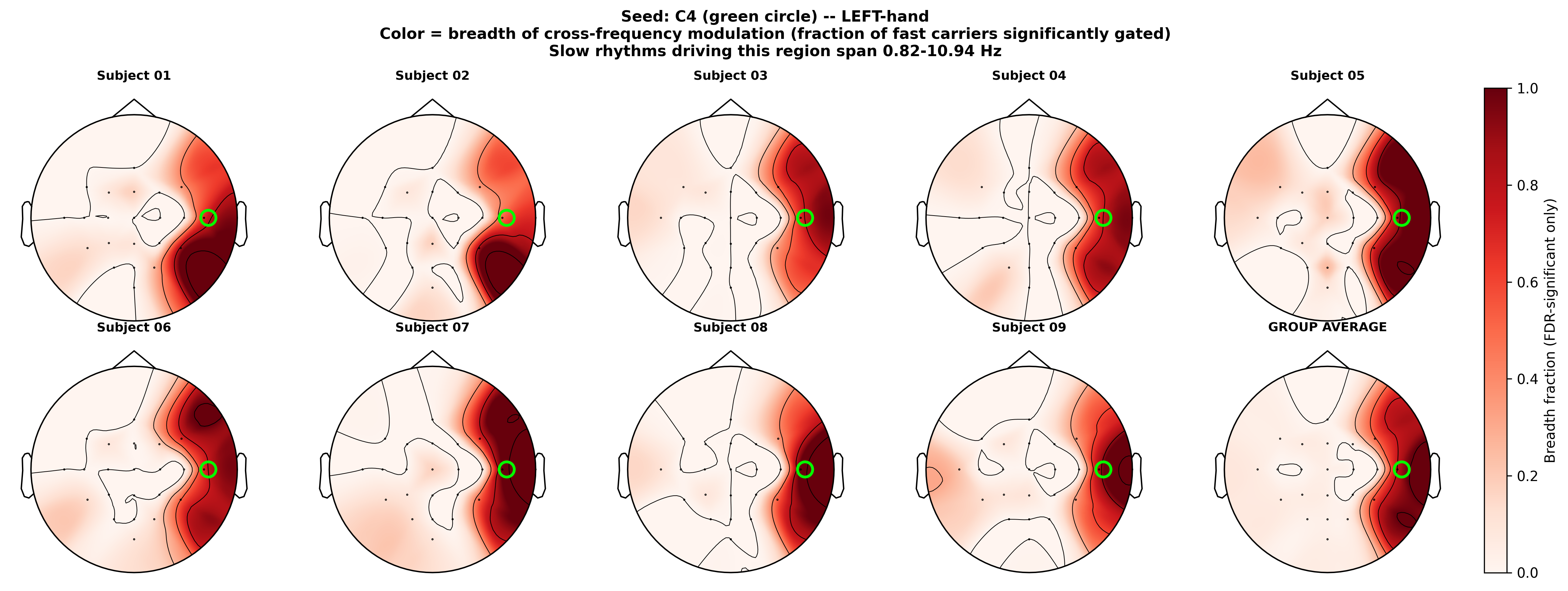} \\
    \vspace{2pt}
    {\small (a) Topographic distribution of $S_2^{\text{norm}}$ breadth fraction for the C4 seed during left-hand motor imagery} \\ \vspace{15pt}
    \includegraphics[width=0.85\textwidth,height=3.5cm,keepaspectratio]
    {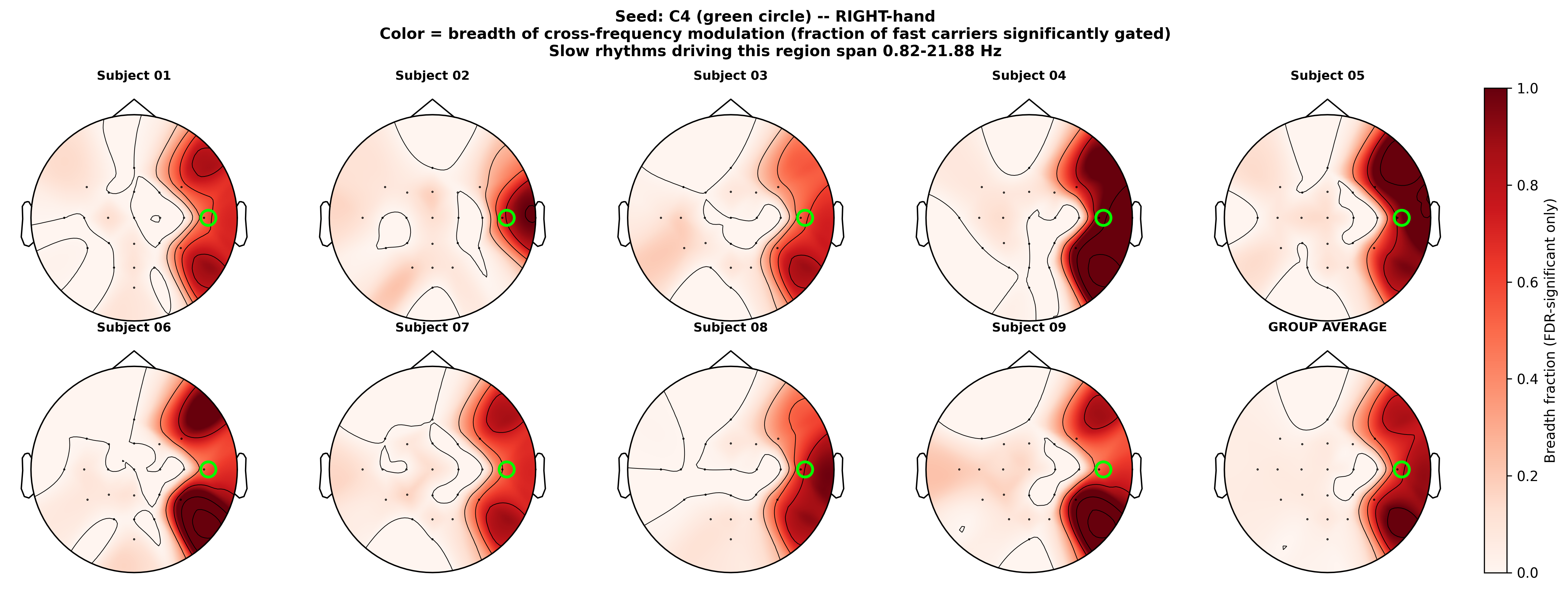} \\
    \vspace{2pt}
    {\small (b) Topographic distribution of $S_2^{\text{norm}}$ breadth fraction for the C4 seed during right-hand motor imagery} \\ \vspace{15pt}
    \includegraphics[width=0.85\textwidth,height=3.5cm,keepaspectratio]
    {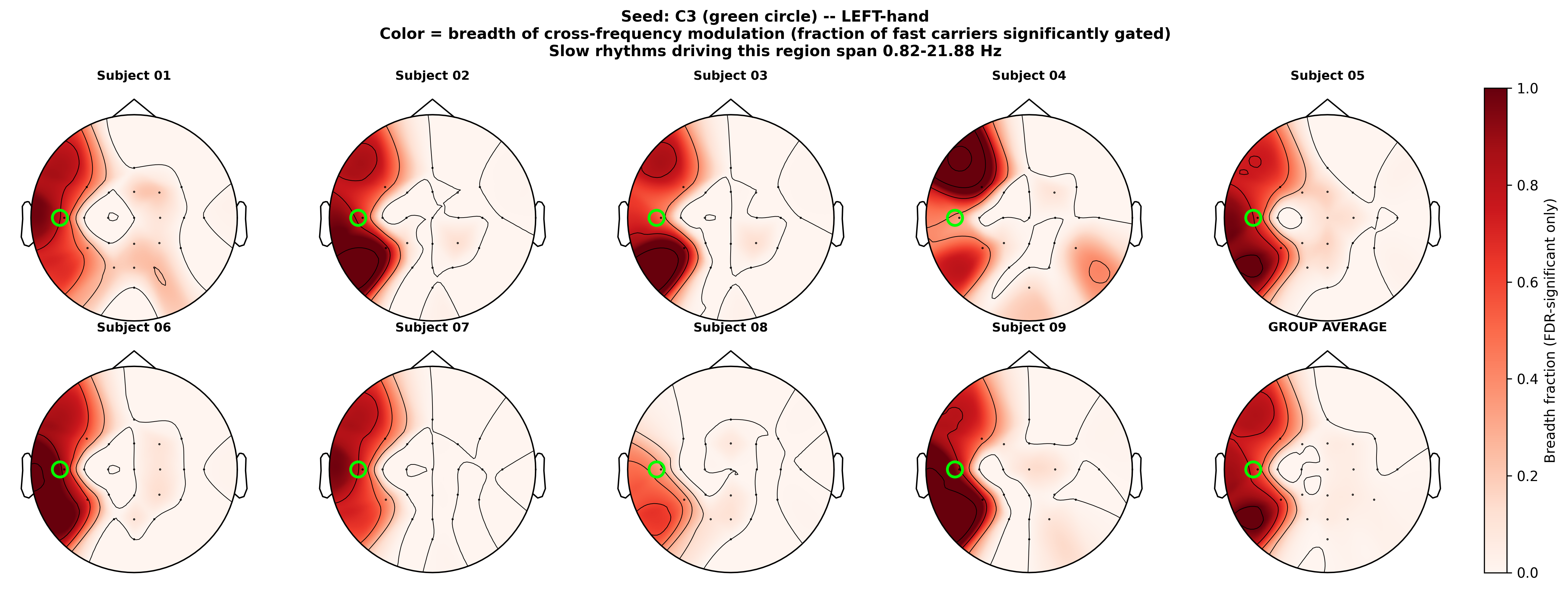} \\
    \vspace{2pt}
    {\small (c) Topographic distribution of $S_2^{\text{norm}}$ breadth fraction for the C3 seed during left-hand motor imagery} \\ \vspace{15pt}
    \includegraphics[width=0.85\textwidth,height=3.5cm,keepaspectratio]
    {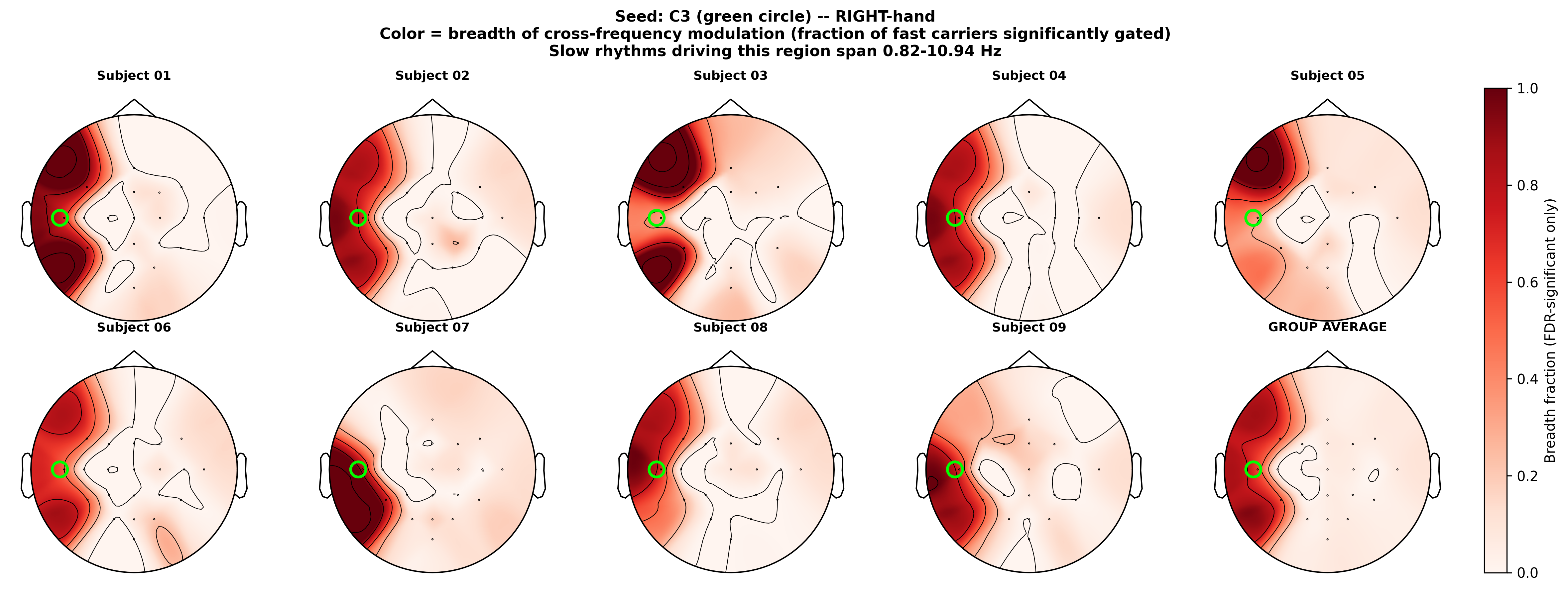} \\
    \vspace{2pt}
    {\small (d) Topographic distribution of $S_2^{\text{norm}}$ breadth fraction for the C3 seed during right-hand motor imagery} \\ \vspace{15pt}
    \caption{eed-based scalp topographies of the breadth-of-modulation fraction ($S_2^{\text{norm}}$) under left- and right-hand motor imagery conditions. Panels demonstrate spatial co-localization of slow-rhythm gating patterns within localized centro-parietal neighborhoods for both C4 and C3 seed configurations.}
    \label{fig:Fig_5}
\end{figure*}

Seed-based topographies of breadth fraction again concentrated within the 
centro-parietal neighborhood identified by $S_1^{\text{norm}}$: for the 
C4 seed, FC4, C6, and CP4 reached substantially higher mean breadth 
($0.48$--$0.76$) than the remaining significant electrodes 
($0.02$--$0.08$), in both conditions. An analogous pattern held for the 
C3 seed (C5, CP3, and FC3 reaching $0.50$--$0.61$) and the Cz seed (C1 
and C2 reaching $0.40$--$0.44$), confirming that the slow-rhythm gating 
structure identified by $S_2^{\text{norm}}$ is spatially co-localized 
with the amplitude-envelope coupling detected by $S_1^{\text{norm}}$, 
rather than distributed uniformly across the scalp.
Seed-based topographies of breadth fraction again concentrated within the centro-parietal neighborhood identified by $S_1^{\text{norm}}$, as illustrated in Figure~\ref{fig:Fig_5}.

\subsection{Phase-Based Connectivity Measures Recover Little of This Structure}

\begin{figure*}[t!]
    \centering
    \includegraphics[width=\textwidth]{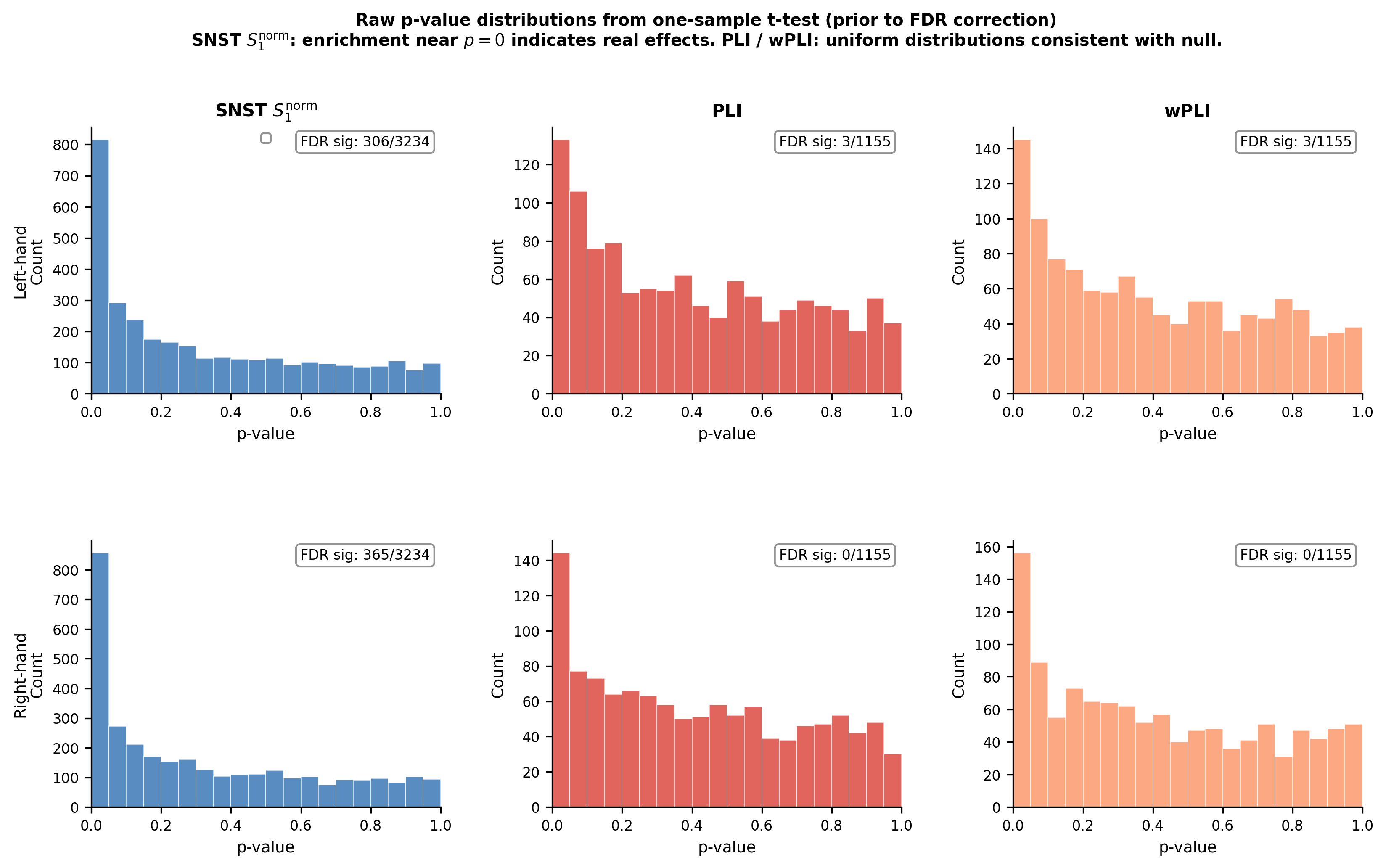}
    \caption{Histograms of raw $p$-value distributions from one-sample $t$-tests (prior to FDR correction)  comparing the proposed SNST $S_1^{\text{norm}}$, PLI, and wPLI metrics.}
    \label{fig:Fig_14}
\end{figure*}
The phase lag index (PLI) and its weighted extension (wPLI) were 
computed from the same epochs and tested under the bias-correction 
and FDR-control procedure. Both measures 
identified markedly fewer significant combinations than 
$S_1^{\text{norm}}$: PLI yielded 3 of 1,155 tested (pair, band) 
combinations for left-hand imagery (0.26\%) and 0 of 1,155 for 
right-hand imagery; wPLI yielded the same counts, with the three 
surviving left-hand combinations identical in pair and effect size 
between the two measures (C2--CP1, mean $z = -0.94$; C6--CP3, mean 
$z = -0.88$; FC1--C1, mean $z = 1.80$; all $p < 0.0001$, $9/9$ sign 
agreement). None of these three combinations involved the C4, C3, or 
Cz seed electrodes; at all three seeds, neither PLI nor wPLI 
identified a single significant target electrode in either condition. 
The organization of inter-regional coupling within EEG motor networks 
has been shown to carry meaningful information about functional 
connectivity, with directed measures revealing distinct roles for 
pre-motor and primary motor areas in sensorimotor network organization 
\cite{b38}, underscoring that the near-total absence of phase-based 
significant coupling found here reflects a genuine methodological 
distinction rather than an absence of inter-regional structure in 
the data.

To compare $S_1^{\text{norm}}$ directly against PLI and wPLI on a common 
basis, each of $S_1^{\text{norm}}$'s significant per-frequency results 
was mapped to the frequency band it falls within, yielding 196 
(left-hand) and 235 (right-hand) distinct significant (pair, band) 
combinations (Table~\ref{tab:sig_combinations}). Of these, zero 
combinations were exclusive to PLI or wPLI, and zero were significant 
across all three measures jointly; 196 (left-hand) and 235 (right-hand) 
combinations were attributable to $S_1^{\text{norm}}$ alone. The statistical robustness of the proposed method is validated by the raw $p$-value distributions shown in Figure~\ref{fig:Fig_14}.

\begin{table*}[t]
\centering
\caption{Significant (pair, band) combinations identified by 
$S_1^{\text{norm}}$ (band-relabeled for comparison), PLI, and wPLI, 
with pairwise overlap.}
\label{tab:sig_combinations}
\begin{tabular}{lccccccc}
\hline
Condition & $S_1^{\text{norm}}$ sig & PLI sig & wPLI sig & 
$S_1^{\text{norm}}$-only & PLI-only & wPLI-only & All three \\ \hline
Left-hand  & 196 & 3 & 3 & 196 & 0 & 0 & 0 \\
Right-hand & 235 & 0 & 0 & 235 & 0 & 0 & 0 \\ \hline
\end{tabular}
\end{table*}

\section{Discussion}\label{sec12}

This study introduced SNST to address a gap recognized across the 
connectivity literature: phase-based measures discard amplitude 
information by construction \cite{b11,b12}, while amplitude-sensitive 
representations, including the wavelet scattering transform, have so far 
been confined to single-channel analysis, as in existing EEG applications 
to seizure detection, schizophrenia classification, and alcoholism 
classification \cite{b15,b16,b42}. The empirical divergence reported 
here---$S_1^{\text{norm}}$ identifying 196--235 robust combinations 
against PLI and wPLI's at most three under an identical correction 
procedure---parallels an independently established finding: amplitude- 
and phase-coupling are dissociable, non-redundant signals reflecting 
at least partly distinct neuronal mechanisms \cite{b26}. This 
dissociation was first demonstrated at the network scale by Hipp et~al., 
who showed that amplitude correlation of band-limited oscillations reveals 
a frequency-specific, spatially organized coupling structure corresponding 
closely to networks identified by functional MRI \cite{b36}, and was 
subsequently formalized as one of two distinct intrinsic coupling modes 
differing in dynamics, origins, and putative function \cite{b37}. A 
recent simulation-based benchmark confirms that amplitude- and 
phase-based metrics carry materially different reliability and artifact 
profiles, such that no single metric is uniformly superior across 
connectivity regimes \cite{b43}. The near-total absence of overlap 
between $S_1^{\text{norm}}$ and the phase-based measures indicates that 
the two families are sensitive to substantially different, largely 
non-redundant aspects of the same signal, not that one measure is 
superior \cite{b26,b37,b43}.

The spatial organization of $S_1^{\text{norm}}$'s significant 
coupling---concentrated at C4--CP4 and its immediate centro-parietal 
neighbors, recurring in both imagery conditions---extends two independent 
findings to a finer spatial scale. Directed connectivity analyses of 
motor-imagery EEG report increased coupling between hemispheres organized 
around the central sensorimotor strip during imagined hand movement 
\cite{b17}, and amplitude correlation specifically recovers spatial 
coupling patterns corresponding to networks established by independent 
imaging modalities \cite{b36}. The present finding extends the 
amplitude-coupling principle from the across-hemisphere and whole-network 
scale to the immediate electrode neighborhood engaged by motor imagery.

The slow-rhythm gating identified by $S_2^{\text{norm}}$---concentrated 
below 11~Hz, dominated by frequencies near 0.82--5.47~Hz---is consistent 
with the established principle that slow amplitude co-fluctuations 
constitute a key mechanism constraining and gating faster network 
activity \cite{b37}. This range additionally overlaps with slow cortical 
potentials, low-frequency EEG shifts indexing cortical excitability 
during motor preparation and execution \cite{b39}, including the 
Bereitschaftspotential, a comparably slow (0--5~Hz) shift preceding 
voluntary movement with peak amplitude organized contralateral to the 
moving limb \cite{b40}. Two independent literatures---amplitude coupling 
as a gating mechanism and slow-potential indices of motor 
excitability---converge on the frequency range identified here by 
$S_2^{\text{norm}}$, suggesting this cross-frequency gating may reflect 
an electrophysiological process already characterized by these methods; 
confirming this directly would require simultaneous recording of both 
signal types.

The absence of comparable findings from PLI and wPLI follows from their 
construction: both discount near-zero-lag phase relationships to suppress 
volume-conduction artifacts \cite{b11,b12}, a trade-off that reduces 
sensitivity to genuine zero-lag and amplitude-only interactions 
\cite{b26,b41}. This robustness is itself incomplete: simulation studies 
show that zero-lag-excluding measures can still yield spurious 
interactions under realistic source-mixing conditions, indicating no 
connectivity measure is immune to every form of estimation bias 
\cite{b41}. SNST's contribution is therefore not a replacement for 
phase-based measures but a recovery of the amplitude-domain portion of 
the connectivity space these measures are constructed to discard.

The centro-parietal regional pattern was identified in the BCI Competition IV-2a dataset alone, indicating this specific regional pattern, unlike the broader $S_2^{\text{norm}}$ slow-rhythm phenomenon, is not yet established as independent of electrode density. Statistical strength is concentrated in a small subset centered on C4--CP4; the 
remaining significant combinations should not each be treated as independently confirmed discoveries given the spatial correlation inherent to dense EEG montages \cite{b6}. The Surface Laplacian applied during preprocessing substantially reduced inter-channel correlation but does not eliminate spatial dependence between adjacent pairs \cite{b21,b32}, and residual correlation may inflate the count of independently significant findings. Moreover, the physiological 
interpretation offered for $S_2^{\text{norm}}$ is indirect, since this study analyzed frequency-domain coupling rather than directly recorded 
slow cortical potentials; confirming this correspondence would require 
simultaneous recording of both. Finally, this work establishes the 
statistical and spatial properties of SNST's descriptors in an offline, two-class analysis; evaluating their utility for online classification, already pursued for amplitude-based connectivity measures and single-channel scattering-transform features \cite{b15,b16,b42}, remains for future work.

\section{Conclusion}\label{sec13}

SNST extends the wavelet scattering transform to the multichannel setting, yielding two complementary descriptors that jointly capture amplitude-envelope coupling between channels and its modulation across frequency scales a connectivity domain structurally. Validated on a publicly available motor-imagery EEG dataset. A direct comparison against the phase lag index and its weighted extension, computed under an identical bias-corrected and false-discovery-rate-controlled pipeline independently verified on synthetic data, confirmed that amplitude-envelope coupling constitutes a largely distinct connectivity signal with negligible overlap with phase-based findings. These results establish SNST as a statistically rigorous, methodologically principled framework for recovering amplitude-domain EEG connectivity structure that existing phase-synchronization approaches are architecturally constrained to discard, with direct applicability to brain-computer interface research and broader multichannel biomedical signal analysis where inter-regional amplitude dependence is of diagnostic or functional interest.

\section*{Declaration of Competing Interest}
The authors declare that they have no known competing financial interests or personal relationships that could have appeared to influence the work reported in this paper.

\section*{CRediT authorship contribution statement}
\textbf{Md. Taksimul Ahsan Tawhid:} Conceptualization, Methodology, Software, Formal analysis, Writing - original draft, Investigation, Project administration. \\
\textbf{Nasif Ahmed Rafe:} Conceptualization, Methodology, Investigation, Validation, Resources, Visualization. \\
\textbf{Alif Tahmid Priyom:} Data curation, Visualization, Resources, Validation.
\\
\textbf{ K. M. Mustafizur Rahman
:} Writing - Review \& Editing, Visualization, Resources, Supervision, Project administration.

\section*{Data Availability Statement}
Validation of the proposed method was conducted using a public four-class motor imagery dataset comprising nine subjects who executed imagination tasks for the left hand, right hand, both feet, and tongue across two distinct recording sessions. The data consists of 22 monopolar EEG channels and 3 EOG channels sampled at 250 Hz, providing 288 trials per subject session alongside expert-scored artifact markers for performance evaluation. link: \url{https://www.bbci.de/competition/iv/#dataset2a}.

\section*{Funding}
This research did not receive any specific grant from funding agencies in the public, commercial, or not-for-profit sectors.

.


\end{document}